\documentstyle[12pt,graphicx,subfigure]{article}
\def\Journal#1#2#3#4{{#1} {\bf #2}, #3 (#4)}

\def\NPB{{\em Nucl. Phys.} B}
\def\PLB{{\em Phys. Lett.}  B}
\def\PRL{\em Phys. Rev. Lett.}
\def\PRD{{\em Phys. Rev.} D}

\def\beq{\begin{equation}}
\def\eeq{\end{equation}}

\def\gmin2{(g-2)_\mu}

\catcode`\@=11 

\def\lsim{\mathrel{\mathpalette\@versim<}}
\def\gsim{\mathrel{\mathpalette\@versim>}}
\def\@versim#1#2{\vcenter{\offinterlineskip
    \ialign{$\m@th#1\hfil##\hfil$\crcr#2\crcr\sim\crcr } }}

\def\PRL{Phys. Rev. Lett.}

\def\beq{\begin{equation}}
\def\eeq{\end{equation}}
\def\beqn{\begin{eqnarray}}
\def\eeqn{\end{eqnarray}}
\def\Journal#1#2#3#4{{#1} {\bf #2}, #3 (#4)}

\def\NPB{{\em Nucl. Phys.} B}
\def\PLB{{\em Phys. Lett.}  B}
\def\PRL{\em Phys. Rev. Lett.}
\def\PRD{{\em Phys. Rev.} D}

\begin{document}

\begin{titlepage}

\begin{center}
{\large {\bf CP Violation Effects on $B^0_{s,d}\rightarrow \it l^+\it l^-$ 
in Supersymmetry\\ at Large $\tan\beta$ }}\\
\vskip 0.5 true cm
\vspace{2cm}
\renewcommand{\thefootnote}
{\fnsymbol{footnote}}
 Tarek Ibrahim$^{a,b}$ and Pran Nath$^{b}$  
\vskip 0.5 true cm
\end{center}

\noindent
{a. Department of  Physics, Faculty of Science,
University of Alexandria,}\\
{ Alexandria, Egypt\footnote{: Permanent address of T.I.}}\\ 
{b. Department of Physics, Northeastern University,
Boston, MA 02115-5000, USA} \\
\vskip 1.0 true cm
\centerline{~ Abstract}
\medskip
An analytic analysis of the CP violating effects arising from the
soft SUSY breaking parameters on the decays 
$B^0_{s,d}\rightarrow \it l^+\it l^-$ at large $\tan\beta$
is given.  It is found that the phases
have a strong effect on the branching ratio and in some regions
of the parameter space they can lead to a variation of the branching
ratio by as much as 1-2 orders of magnitude. These results
have important implications for  the discovery of  
the $B^0_{s}\rightarrow \mu^+\mu^-$ signal in RUNII of the
Tevatron and further on how the
parameter space of SUSY models will be limited once the signal
is found. 
\end{titlepage}
\section{Introduction}
Recently there has been a great amount of interest in the rare process
$B^0_s\rightarrow \mu^+\mu^-$\cite{Babu:1999hn,Bobeth:2001sq,Chankowski:uz}
as it offers an opportunity to probe physics  beyond
the standard
model\cite{Babu:1999hn,Bobeth:2001sq,Chankowski:uz,Dedes:2002zx,Arnowitt:2002cq,baek,tata}.
Thus in the standard model the branching ratio is
 rather small\cite{Bobeth:2001sq}, i.e., 
 $ B(\bar B^0_s\rightarrow \mu^+\mu^-)$
 $=$$(3.1\pm 1.4)\times 10^{-9}$ (for $|V_{ts}|=0.04\pm 0.002$)
 while in supersymmetric models it can get three orders of magnitude
 larger  for large $\tan\beta$ and the branching ratio
  can be as large as $10^{-6}$. 
 This result is very exciting in view of the fact that the 
 sensitivity of the Tevatron to this decay will improve by two 
 orders of magnitude allowing a test of a class of supersymmetric 
 models even before any sparticles are found. 
 Thus while the current limit on this decay is 
 $ B(\bar B^0_s\rightarrow \mu^+\mu^-)$
 $<$$(2.6)\times 10^{-6}$ it is estimated that the RUNII of the
 Tevatron could be sensitive to a branching ratio down to the
 level of $10^{-8}$ or even lower\cite{Arnowitt:2002cq}. 
While this sensitivity is still too small
 to test the standard model prediction, it is large enough to 
 explore significant portions of the parameter space of supersymmetric
 models such as mSUGRA\cite{msugra}. 
The previous analyses have mostly been in the context of CP conservation
except for the works of Refs.\cite{randall,huang}. In Ref.\cite{randall}
effects of CP violation on lepton asymmetries, ie., 
$(l^+l^+-l^-l^-)/ (l^+l^++l^-l^-)$
in decays of $B\bar B$ pairs were studied, while in Ref.\cite{huang}
CP asymmetry in B and $\bar B$ decays was investigated but only a 
cursory mention of the effect of CP violation on the size of the branching
ratio, which is primarily the quantity which will be measured at the
Tevatron, was made in Ref.\cite{huang}. The effects of CP phases on the
 Higgs sector were ignored in these works. However, it is known that
 at large $\tan\beta$ CP mixing effects in the neutral Higgs sector
 are very significant and cannot be ignored\cite{pilaftsis,inhiggs1}. 
 In this analysis we give
 a complete analysis of the effects of CP violation on the decay
 $B^0_{d,s}\rightarrow \it l^+\it l^-$  valid at
 large $\tan\beta$ including the effects of CP violation in the
 neutral Higgs sector which mediates the decay. The focus of our
 work is the effect of CP violation on the branching ratio
 $ B(\bar B^0_{d,s}\rightarrow \it l^+\it l^-)$.
 Specifically we would like to see the size of the variation in the
 branching ratio when the phases  are included in the analysis 
 and to see if such variations will allow the branching ratio to lie
 within reach of RUNII of the Tevatron.

In supersymmetric models
 CP phases arise naturally via the soft breaking masses.
 Thus the mSUGRA model allowing for  complex soft parameters
  contains two phases and the parameter space of
 such models can be characterized by $m_0,m_{\frac{1}{2}}, |A_0|, \tan\beta,
 \theta_{\mu}, \alpha_{A}$ where $m_0$ is the universal scalar mass,
 $m_{\frac{1}{2}}$ is the universal gaugino mass, $|A_0|$ is the universal
 trilinear coupling, $\tan\beta=<H_2>/<H_1>$ where $H_2$ gives mass to 
 the up quark and $H_1$ gives mass to the down quark and the leptons,
 $\theta_{\mu}$ is the phase of the Higgs mixing parameter $\mu$ and
 $\alpha_{A}$ is the phase of the trilinear couplings $A_0$.
 In extended SUGRA models with  nonuniversalities and in the minimal
 supersymmetric standard model  (MSSM) one can have many more phases. 
 Specifically the $U(1), SU(2), SU(3)$  gaugino masses $m_i$ (i=1,2,3)
 can have phases so that $m_i=|m_i| e^{i\xi_i}$ and such phases
 play an important role in SUSY phenomena at low energy.
 An important constraint on models  with CP violation is that of the
 experimental limits on the electron and on the neutron electric dipole
 moments (edms)($d_e< 4.3\times 10^{-27}ecm$\cite{eedm},
 $ d_n<6.5\times 10^{-26} ~ecm$\cite{nedm}).
 These constraints can be satisfied in a variety  
 of ways\cite{ellis,na,bdm2,incancel,chang}.
 The limit on the edm of $H_g^{199}$ is also known to a high degree of 
 accuracy ($d_{H_g}<9\times 10^{-28} ecm$\cite{hg199}) and recent 
 analyses have also included this constraint\cite{olive}. 
  Specifically in scenarios with the cancellation 
  mechanism\cite{incancel}
   and in scenario
 with phases only in the third generation\cite{chang} one can accommodate large CP
 violating phases and  their inclusion can affect supersymmetric phenomena
 in a very significant way. 
 In this work we will focus on the contribution from the so called counter term 
 diagram (Fig.1) which gives an amplitude proportional to $\tan^3\beta$
 for large $\tan\beta$. 
The decay  $B^0_{d'}\rightarrow \it l^+\it l^-$  ($d'=d,s$) is governed  by the 
effective Hamiltonian\cite{Bobeth:2001sq} 
\begin{eqnarray}
H_{eff}=-\frac{G_Fe^2}{4\sqrt 2 \pi^2} V_{tb}V_{td'}^* 
(C_S O_S + C_P O_P +C_S' O_S'\nonumber\\ 
+ C_P' O_P'+ + C_{10} O_{10})_Q
\end{eqnarray}
where 
\begin{eqnarray}
O_S= m_b  (\bar d'_{\alpha}P_Rb_{\alpha})(\bar l l),~~
O_P=m_b (\bar d'_{\alpha}P_Rb_{\alpha})(\bar l\gamma_5 l),\nonumber\\
O_S'= m_{d'}  (\bar d'_{\alpha}P_Lb_{\alpha})(\bar l l),
O_P'= m_{d'}  (\bar d'_{\alpha}P_Lb_{\alpha})(\bar l \gamma_5 l),\nonumber\\
O_{10}=  (\bar d'_{\alpha}\gamma^{\mu}P_Lb_{\alpha})
(\bar l\gamma_{\mu}\gamma_5 l)
\end{eqnarray}
and where the subscript $Q$ in Eq.(1) is the scale where the
 quantities are being evaluated.
The branching ratio $B(B^0_{d'}\rightarrow \it l^+\it l^-)$
 is given by
  \beqn
 B(\bar B^0_{d'}\rightarrow \it l^+\it l^-) =
 \frac{G_F^2\alpha^2M_{B_{d'}}^5\tau_{B_{d'}}}{16\pi^3}
 |V_{tb}V_{td'}^*|^2 \nonumber\\
  (1-\frac{4 m_{\it l}^2}{M^2_{B_{d'}}})^{1/2}
 \{(1-\frac{4 m_{\it l}^2}{M^2_{B_{d'}}}) |f_S|^2+|f_P+2m_{\it l} f_A|^2\}
 \eeqn
 where $f_i$ (i=S,P) and $f_A$ are defined as  follows
 \beq
 f_i= -\frac{i}{2}f_{B_{d'}}(\frac{C_im_b-C_i'm_{d'}}{m_{d'}+m_b}), 
 f_A=-\frac{if_{B_{d'}}}{2M_{B_{d'}}^2} C_{10}
 \eeq
  Additionally in the above one
should include the SUSY QCD correction\cite{carena}
 which behaves like $\mu\tan\beta$ and can produce
a significant effect in the large $\tan\beta$ region.

\begin{figure}
\hspace*{-0.6in}
\centering
\includegraphics[width=9cm,height=6cm]{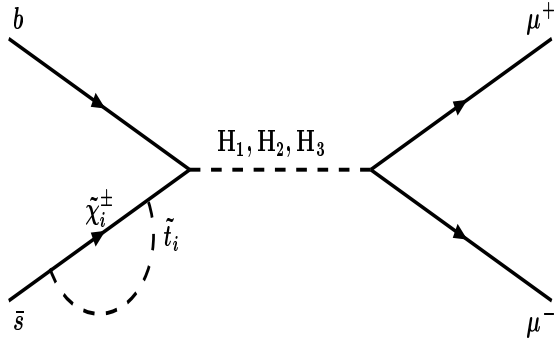}
\caption{The counter term diagram.
 which contributes to the branching ratio
$B^0_{s}\rightarrow \it l^+\it l^-$ and produces the leading term in the 
amplitude proportional to $\tan^3\beta$. }
\label{1}
\end{figure}
\section{Effects of CP Violation on $B_{s}^0\rightarrow \mu^+\mu^-$}
We discuss now the effects of CP violation on the decay 
$B_{s}^0\rightarrow \mu^+\mu^-$.
As pointed out above the diagram which gives the largest contribution
in the large $\tan\beta$ region is the counter term diagram and 
involves the exchange of Higgs poles. 
In the absence of 
CP violation the Higgs sector diagonalizes into a CP even and 
a CP odd sector. However,
it is well known that loop effects induce a  CP violation in the
Higgs sector and generate a mixing of CP even and CP 
odd sectors\cite{pilaftsis,inhiggs1}.
As a consequence CP effects will enter Fig.1 not only in the vertices
involving charginos and squarks but also in the Higgs poles
and the vertices involving the Higgs. 
CP violation in the Higgs sector
can be exhibited by parameterizing the Higgs VEVs in the presence
of CP violating phases as follows: 
\beqn
(H_1)=(H_1^0, H_1^-)=(v_1+\phi_1+i\psi_1, H_1^-)/\sqrt 2,\nonumber\\
(H_2)= (H_2^+, H_2^0) =e^{i\theta_H} (H_2^+,
             v_2+\phi_2+i\psi_2)/\sqrt 2)
\eeqn	      
where in general $\theta_H$ is non-vanishing as a consequence of the 
minimization conditions of the Higgs potential. 
In the presence of CP
 violating phases
the CP even and the CP odd sectors of the Higgs fields mix 
and thus the Higgs matrix is a $4\times 4$ matrix in the basis 
$\phi_1, \phi_2, \psi_1, \psi_2$. 
In the basis
$\phi_1, \phi_2, \psi_{1D},\psi_{2D}$ where 
\begin{eqnarray}
\psi_{1D}=\sin\beta \psi_1+ \cos\beta \psi_2,\nonumber\\ 
\psi_{2D}=-\cos\beta \psi_1+\sin\beta \psi_2
\end{eqnarray}
the  $\psi_{2D}$ field decouples and is identified as the Goldstone and
one is left with a remaining $3\times 3$  $(mass)^2$ matrix 
which mixes CP even and CP odd states. 
The Higgs mass matrix can be diagonalized by the transformation
 \begin{eqnarray}
R M^2_{Higgs} R^T =diag (M_{H_1}^2, M_{H_2}^2, M_{H_3}^2)
\end{eqnarray}
where the eigen values $(m_{H_1}, m_{H_2},m_{H_3})$ are now
admixtures of CP even and CP odd states and we arrange the
eigenvalues so that in the limit of no CP violation 
one has the identification $(m_{H_1}, m_{H_2},m_{H_3})$$\rightarrow$ 
$(m_H, m_h, m_A)$ where (h, H) are the (light, heavy) CP even Higgs
bosons and A is the CP odd  Higgs boson. 
The diagonalization modifies the 
vertices which connect the Higgs with the quarks and the leptons.
One finds that the interaction Lagrangian for the Higgs vertices that
enter in the counter term diagram is now given by
\begin{eqnarray}
L_{ffH}=-(g m_b/2 m_W cos \beta)\Sigma_{j=1}^3\bar{b}[R_{j1}-
i\gamma_5 sin\beta  R_{j3}]b H_j\nonumber\\
-(g m_l/2 m_W cos \beta)\Sigma_{j=1}^3 \bar{l} 
[R_{j1}-i \gamma_5 sin\beta R_{j3}] l H_j 
\end{eqnarray}
In the SUSY sector we carry out the analysis in the 
scenario with minimal flavor violation where the squark mass 
matrices are assumed flavor-diagonal. Under this assumption 
we find
\beqn
C_S=-\frac{m_{\it l}}{\sqrt 2 m_W^2 \cos^3\beta} 
\sum_{j=1}^{3}\sum_{s=1}^{2} 
m_{\chi_s^+}\frac{R_{j1}^2}{M_{H_j}^2} \psi_s
\eeqn
\beqn
C_P=\frac{m_{\it l}\tan^2\beta}{\sqrt 2 m_W^2 \cos\beta} 
\sum_{j=1}^{3}\sum_{s=1}^{2} 
m_{\chi_s^+}\frac{R_{j3}^2}{M_{H_j}^2} \psi_s
\eeqn
\beqn
\psi_s=\sum_{q=t,c,u} U_{s2}\lambda_{qq}\{(\frac{m_W V_{s1}}{2\sin^2\theta_W})
[ \cos^2{\frac{\theta_q}{2}}  
f_3({m^2_{\tilde q_1}}/{m_{\chi_s^{+2}}})\nonumber\\
+ \sin^2{\frac{\theta_q}{2}}  f_3({m^2_{\tilde q_2}}/{m_{\chi_s^{+2}}})]
+\frac{m_qV_{s2} \sin\theta_q}{4\sqrt 2 \sin^2\theta_W \sin\beta} 
e^{i\phi_q}\Delta f_3\}
\eeqn
where $\Delta f_3$= $f_3({m^2_{\tilde q_2}}/{m_{\chi_s^{+2}}})$-
$f_3({m^2_{\tilde q_1}}/{m_{\chi_s^{+2}}})$, 
$\tilde q_1(\tilde q_2)$ is the heavier (lighter) squark
 and   $\theta_q$ and $\phi_q$ are defined by
\beq
\sin\theta_q=\frac{2m_q|A_q m_0-\mu^*\cot\beta|}
{m_{\tilde q_1}^2-m_{\tilde q_2}^2}
\eeq
\beq
\sin\phi_q=\frac{m_0|A_q|\sin\alpha_{A_q} +|\mu| \sin\theta_{\mu}
\cot\beta}{|m_0A_q -\mu^* \cot\beta|}
\eeq
In Eq.(11) $U$ and $V$ are the diagonalizing matrices for the 
chargino mass matrix $M_C$ so that
$U^* M_C V^{-1} =diag (m_{\chi_1^{\pm}},m_{\chi_2^{\pm}})$
and finally $f_3$ in Eq.(11) is a form factor defined by 
$f_3(x)={x}lnx/(1-x)$. 
For the diagram of Fig.1 $C_S'$ and $C_P'$  are defined as follows
\beqn
C_S'(counter)= (C_S(counter)[\lambda_{qq} \rightarrow 
\lambda_{qq}^*])^*\nonumber\\
C_P'(counter)= -(C_P(counter)[\lambda_{qq} \rightarrow  
\lambda_{qq}^*])^*
\eeqn
where $\lambda_{qq}=(V_{qb}V_{qd'}^*/V_{tb}V_{td'}^*)_{CKM}$ 
  and the subscript means that the matrix V here  is the CKM matrix. 
Eqs.(9-13) give the most general result for the minimal flavor violation  
with inclusion of phases without any approximations. 
 Neglecting the squark mixings for the first two generations Eq.(11)
 simplifies so that 
\beqn
\psi_s=U_{s2}\{(\frac{m_W V_{s1}}{2\sin^2\theta_W})
[\lambda_{uu} f_3(y_{\tilde u_{1s}}) 
+ \lambda_{cc} f_3(y_{\tilde c_{1s}})\nonumber\\
+ \cos^2{\frac{\theta_t}{2}} f_3(y_{\tilde t_{1s}})
+\sin^2{\frac{\theta_t}{2}} f_3(y_{\tilde t_{2s}})] \nonumber\\
+\frac{m_tV_{s2} \sin\theta_t}{4\sqrt 2 \sin^2\theta_W \sin\beta} 
e^{i\phi_t} [ f_3(y_{\tilde t_{2s}}) -f_3(y_{\tilde t_{1s}})]\}
\eeqn
where $ y_{\tilde u_{1s}}$ etc are defined by
$y_{\tilde u_{1s}}= ({m^2_{\tilde u_1}}/{m_{\chi_s^{+2}}})$,
$y_{\tilde c_{1s}}= ({m^2_{\tilde c_1}}/{m_{\chi_s^{+2}}})$,
$y_{\tilde t_{1s}}= {m^2_{\tilde t_1}}/{m_{\chi_s^{+2}}}$,
$y_{\tilde t_{2s}}= {m^2_{\tilde t_2}}/{m_{\chi_s^{+2}}}$.

\section{Numerical Size of CP Effects}
Eqs.(9)-(15) constitute the new results 
of this analysis as they include the effects of CP violation.
In the limit when we neglect the CP phases our approximation
 Eq.(15) agrees with
the result of Bobeth etal.\cite{Bobeth:2001sq} at large $\tan\beta$.
 The phases 
 can increase or decrease the branching ratio. We focus on the 
 region of the parameter space where an enhancement occurs and
 this region is of considerable relevance for the detection of the 
 $B^0_s\rightarrow \mu^+\mu^-$ signal.
\begin{figure}
\hspace*{-0.6in}
\centering
\includegraphics[width=9cm,height=6cm]{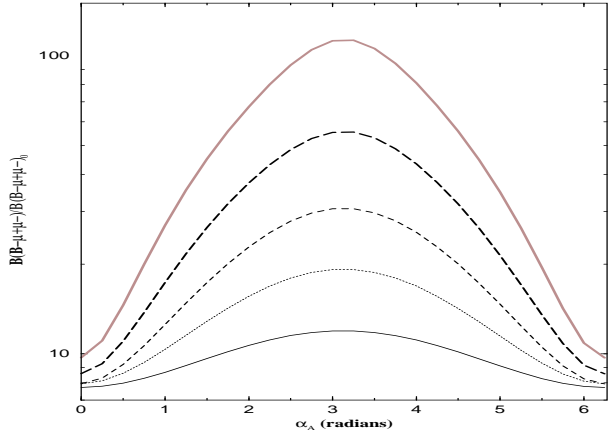}
\caption{The ratio of the branching
ratios $B(B^0_s\rightarrow \mu^+\mu^-)/B(B^0_s\rightarrow \mu^+\mu^-)_0$,
where $B(B^0_s\rightarrow \mu^+\mu^-)_0$ is the branching ratio when all phases
are set to zero,  
as a function of the CP violating
phase $\alpha_A$ for values of $|A_0|$ of 1,2,3,4,5 in the ascending
order of the curves.
The other parameters are $m_0=200$ GeV, $m_{1/2}=200$ GeV, 
 $\tan\beta =50$, $\xi_1=\xi_2=\pi/4$, $\xi_3=0$, and $\theta_{\mu}=2$.}
\label{2}
\end{figure}
\begin{figure}
\hspace*{-0.6in}
\centering
\includegraphics[width=9cm,height=6cm]{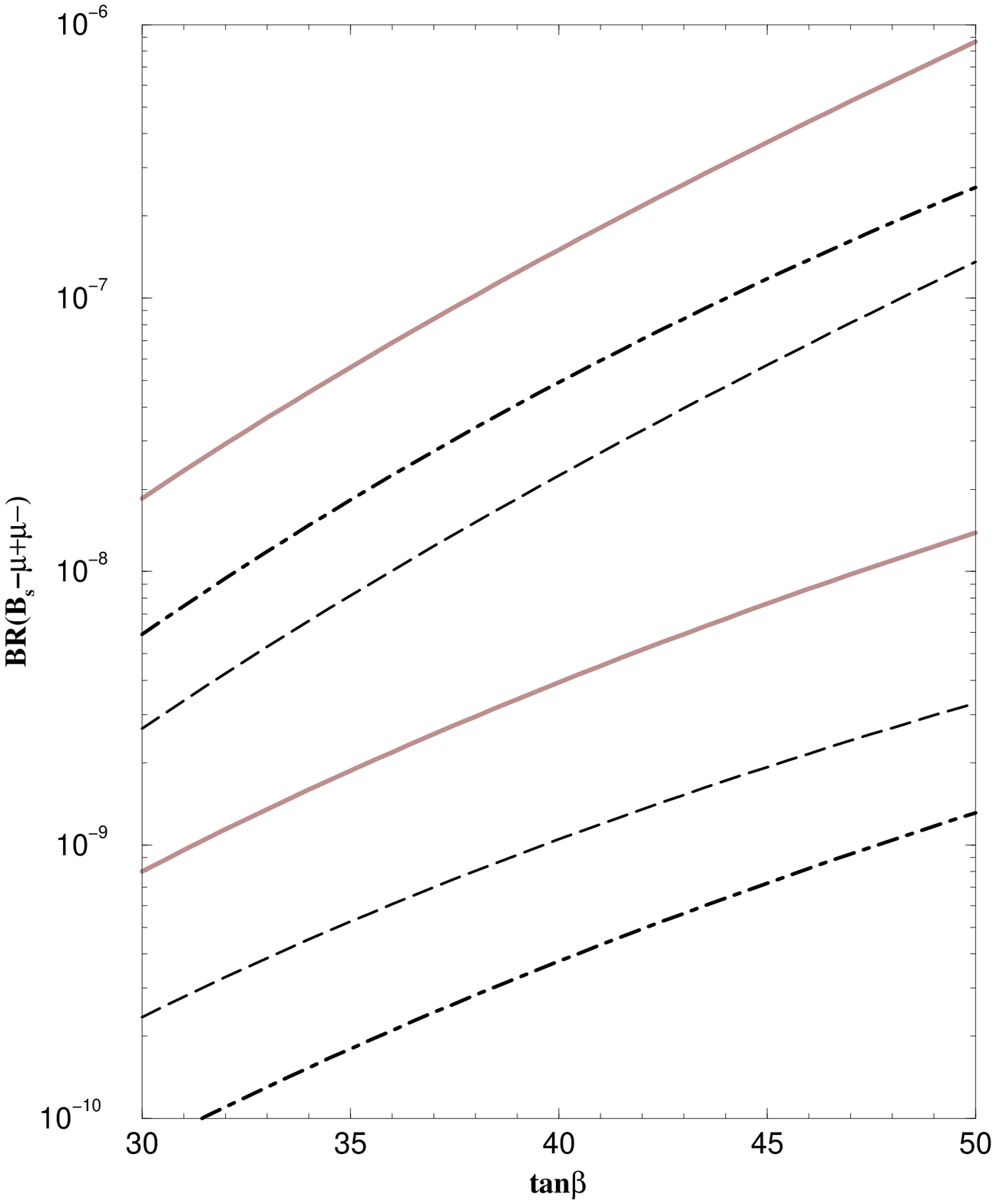}
\caption{Exhibition of the branching ratio $B^0_s\rightarrow \mu^+\mu^-$
 as a function of $\tan\beta$. The three top curves correspond in 
 descending order to cases (a), (b) and (c) of Table 1 where the
 input parameters for each case is recorded (all masses in GeV and 
 all angles in radians). The lower solid, dashed and dot-dashed 
 curves have the same inputs as the top solid, dashed and dot-dashed
 curves except that all phases are set to zero.  The edms corresponding 
 to the top three curves for the case $\tan\beta =50$ are  given in Table1.} 
\label{3}
\end{figure}
\begin{figure}
\hspace*{-0.6in}
\centering
\includegraphics[width=9cm,height=6cm]{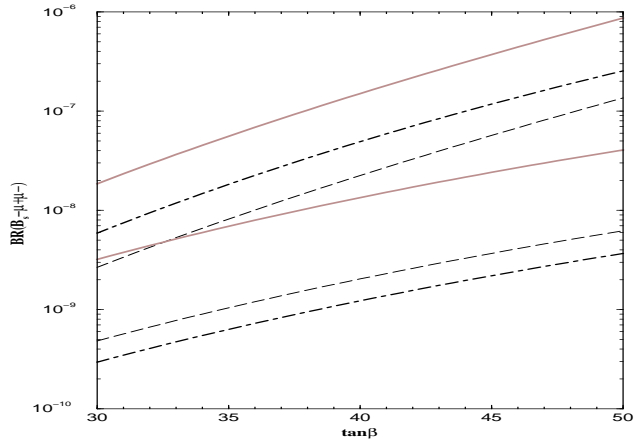}
\caption{Same as  Fig.3 except that the lower three curves
have all phases set to $\pi$.}
\label{4}
\end{figure}
\begin{center} 
\begin{tabular}{|c|c|c|c|c|c|}
\multicolumn{6}{c}{Table 1. Electron, neutron and $H_g$ edms} \\
\hline
\hline
case& $m_0$, $m_{\frac{1}{2}}$, $|A_0|$ & 
$\alpha_A$, $\xi_1$, $\xi_2$, $\xi_3$ &$d_e$ & $d_n$& $d_{H_g}$\\
\hline
\hline
(a) & $200, 200, 4$ & $1, .5, .659, .633$ & $1.45\times 10^{-27}$ 
& $9.2\times  10^{-27}$ &   $7.2\times  10^{-27}$ \\
\hline 
(b) & $370, 370, 4$ & $2, .6, .653, .672$ & $-1.14\times 10^{-27}$ &
$-7.9\times 10^{-27}$ & $2.87\times 10^{-26}$ \\
\hline
(c) & $320, 320, 3$ & $.8, .4, .668, .6$ & $-3.5\times 10^{-27}$ &
$7.1\times 10^{-27}$ &  $2.9\times 10^{-26}$ \\
\hline 
\hline
\end{tabular}
\end{center}
 In Fig.2 we give a plot of the ratio 
 $B(B^0_s\rightarrow \mu+\mu^-)/B(B^0_s\rightarrow \mu+\mu^-)_0$,
 where $B(B^0_s\rightarrow \mu+\mu^-)_0$ is the branching ratio in the
 absence of phases, as a function
 of the phase $\alpha_A$ of the trilinear coupling $A_0$ for values
 of $|A_0|$  ranging from 1 to 5. In each case we find
 that the ratio of  
 branching ratios shows a strong dependence on $\alpha_A$.
 Specifically
  for the case  $|A_0|=5$ the ratio can become as large as $10^2$.
  Thus Fig.(2) shows that with phases the branching ratio can be
  significantly modified. 
   There are many contributing factors
  to this phenomenon. Thus in Eqs.(9),(10) and (15) we find several
  quantities that depend on the phases. These include the
  chargino masses, the Higgs masses, the mixings $R_{j1}$ 
  and $R_{j3}$ etc. However, the largeness of the
  effect arises mostly from the variation in $\psi_s$. 
   Here one finds that in
some regions of the parameter space the masses
of the stops and their mixings are strongly affected by the CP
phases which affect $\sin\theta_t$ and $f_3$ and  their
combined effect can generate a large enhancement of the
amplitude when the phases are included.
In Fig.3 we exhibit the branching ratio $B^0_s\rightarrow \mu^+\mu^-$
as a function of $\tan\beta$ for various values of the phases and 
compare the results to the CP conserving case where the phases 
are all set to zero. The analysis of Fig.4 is identical to that of
Fig.3 except that a comparison is made with the CP conserving case
where the phases are all set to $\pi$. 
One finds that often points in the parameter space which would otherwise
(i.e., when phases all vanish or are equal to $\pi$) lie below the 
sensitivity of $\sim 10^{-8}$ for the branching ratio,
which is what the RUNII of the Tevatron can achieve in the future,
can now be moved into the region of sensitivity  of the Tevatron.

We have checked that using the cancellation mechanism there exist
regions of the parameter space where the experimental constraints
on the edms of the electron, of the neutron, and of $H_g^{199}$
 are satisfied. For $H_g^{199}$ the atomic edm constraint can be  translated
 into a contrainst on a specific combination of the chromo electric dipole
 moment of u, d and s quarks, so that 
 $C_{H_g} =|d_d^C-d_u^C-0.012 d_s^C|$ is contrained  to satisfy
 $C_{H_g}<3.0\times 10^{-26}cm$.
 An example is given in Table1. As shown 
 in Ref.\cite{inscaling} one uses 
scaling to generate a trajectory where cancellations occur and the edm
constraints are satisfied starting from a given cancellation point.
We have checked that this is the case for the region such as in Table1.
In Fig.5 we exhibit the effect of CP phases on the branching ratio of
 $B^0_d\rightarrow \tau^+\tau^- $  and compare the result to the CP
 conserving cases where all the phases are set to zero. 
 (An analysis without phases for this process is also given in 
 Ref.\cite{tata}). The analysis of Fig.6 is identical to that of 
 Fig.6 except that a comparison with CP conserving cases is made 
 by setting all the phases to $\pi$. It would be interesting
  to see if some of the  region of Figs. 5 and 6 would be accessible 
  to experiment in the future. 
  An interesting study has recently appeared 
  correlating the $B^0_{d,s}-\bar B^0_{d,s}$ mass difference 
  $\Delta M_{d,s}$ with $B^0_{d,s}\rightarrow \mu^+\mu^-$ in
  supersymmetry for large $\tan\beta$\cite{buras}. It would be interesting to 
  investigate this correlation in the presence of phases in MSSM.
  However, this requires an analysis of the  $B^0_{d,s}-\bar B^0_{d,s}$ 
  mass difference with inclusion of phases in the supersymmetric
  loop contribution to the $\Delta M_{d,s}$. 
  This possibility is under investigation. 
  
  \section{Conclusion}
In conclusion in this work we have derived analytic results for
the effects of CP violating phases on the branching ratio 
$B^0_{d'}\rightarrow \it l^+\it l^-$ arising from the chargino-stop
exchange contribution in the counter term diagram in the large
$\tan\beta$ region. It is found that the branching ratio in general is
sensitive to the CP phases and that the  branching ratio can vary in
some parts of the parameter space by up to 1-2 orders
of magnitude.
These results have important implications for the search for
this signal and for the interpretation of it in limiting the
SUSY parameter space once the signal is found. 
Of course significant effects of order 50-100\% can be obtained with
significantly smaller phases than those in Table 1 and here the edm
constraints can be satisfied over a much larger parameter space.
In the above analysis we have not included the effects of the 
gluino and the neutralino exchanges. Inclusion of these will bring
in a strong dependence on additional phases $\xi_1$ and $\xi_3$. 
A full analysis of the CP violating effects valid also for
small $\tan\beta$ will be discussed elsewhere.

\begin{figure}
\hspace*{-0.6in}
\centering
\includegraphics[width=9cm,height=6cm]{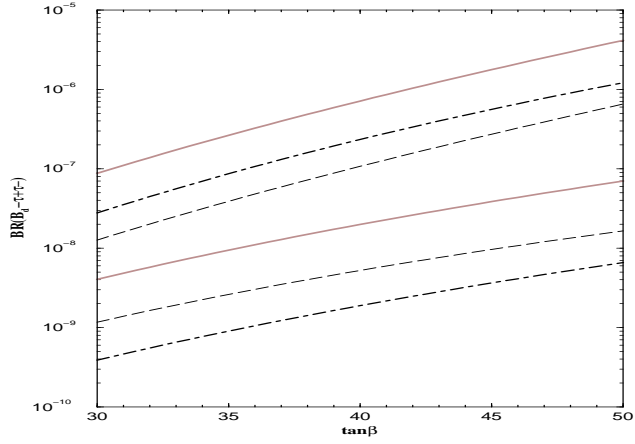}
\caption{Exhibition of the branching ratio
 $B^0_d\rightarrow \tau^+\tau^-$
 as a function of $\tan\beta$. The three top curves correspond in 
 descending order to cases (a), (b) and (c) of Table 1 where the
 input parameters for each case is recorded (all masses in GeV and 
 all angles in radians). The lower solid, dashed and dot-dashed 
 curves have the same inputs as the top solid, dashed and dot-dashed
 curves except that all phases are set to zero.  The edms corresponding 
 to the top three curves for the case $\tan\beta =50$ are  given in Table1.} 
\label{5}
\end{figure}
\begin{figure}
\hspace*{-0.6in}
\centering
\includegraphics[width=9cm,height=6cm]{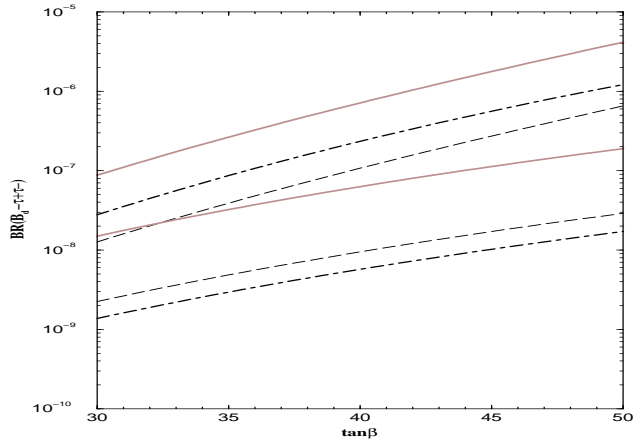}
\caption{Same as Fig.5 except that the lower three curves
have all phases set equal to $\pi$.}  
\label{6}
\end{figure}

\begin{center}
{\bf ACKNOWLEDGEMENTS}
\end{center}
This research was supported in part by NSF grant  PHY-0139967.

\end{document}